\documentstyle[preprint,aps,prb,epsf]{revtex}

\begin{document}

\title{Amino acid classes and the protein folding problem}

\author{Marek Cieplak$^{1,2}$, Neal S. Holter$^1$, Amos Maritan$^3$ and
Jayanth R. Banavar$^1$}
\address{$^1$ Department of Physics and Center for Materials Physics,
104 Davey Laboratory, The Pennsylvania State University, University
Park, Pennsylvania 16802}
\address{$^2$Institute of Physics, Polish Academy of Sciences,
Al. Lotnikow 32/46, 02-668 Warsaw, Poland}
\address{$^3$ International School for Advanced Studies (SISSA),
Via Beirut 2-4, 34014 Trieste, INFM and the Abdus Salam
International Center for Theoretical Physics, Trieste, Italy}

\address{
\centering{
\medskip\em
{}~\\
\begin{minipage}{14cm}
We present and implement a distance-based clustering of amino acids 
within the framework of a statistically derived interaction matrix
and show that the resulting groups faithfully reproduce, for well-designed
sequences, thermodynamic
stability in and kinetic accessibility 
to the native state.
A simple interpretation of the groups is obtained by eigenanalysis
of the interaction matrix.
{}~\\
{}~  
\end{minipage}
}}

\maketitle
\newpage
The principal theme of this paper is to address the issue of determining
the minimum number of distinct amino acids
that are needed to make protein-like sequences with folds similar to those
found in nature.  A general answer to this question would have
important ramifications in the design of proteins
and in the origin of life.
Our analysis is carried out within the framework of
the Miyazawa-Jernigan (MJ) \cite{Miyazawa} 
interaction matrix which was derived by them 
using a statistical approach and is commonly used as a measure of
the coarse-grained interactions between amino acids in a protein.

Experiments \cite{Kauzmann,Kamtekar} 
and theoretical studies \cite{Dill} have shown the critical role played by
two kinds of residues: hydrophobic and polar.  There is an effective attraction
between hydrophobic amino acids that arises from their aversion to
the solvent and lead to such amino acids forming the core 
in the protein native state.
Recent protein engineering experiments suggest that maybe not two
but certainly several amino acids can
effectively substitute for the full 20-alphabet set.
It has been shown \cite{Regan} that helical bundles can be built with 
the set of three amino acids:
hydrophobic L (leucine), polar E (glutamic acid), and polar K (lysine).
Helical bundles have also been found \cite{Kamtekar}
to be viable when hydrophobic sites
are filled from the set [F (phenylalanine), L, I (isoleucine),
M (methionine), V (valine)] and the polar sites
from the set [E, D (aspartic acid), K, N (aspargine), Q (glutamine),
H (histidine)].
Encoding the $\beta$-sheet SH3 domain, however, requires
five amino acids: hydrophobic I (isoleucine), K, E, A (alanine), 
and G (glycine) \cite{Riddle}.
This suggests that the 20 amino acids can be grouped
into five distinct clusters with the members of each group having
quite similar properties.
Riddle et al. \cite{Riddle} and Wolynes \cite{Wolynes}
have presented persuasive arguments that five groups are needed 
to provide enough
specificity to form a folding funnel and generate few traps
in the energy landscape.

Wang and Wang \cite{Wang} have suggested that a justification for
the five group clustering scheme can be provided by minimizing
the mismatch between the reduced and complete interaction matrices.
Specifically, they considered the MJ
\cite{Miyazawa} matrix
and deduced a clustering
scheme, shown at the top of Figure 1, in which the representative 
amino acids were IKEAG -- precisely as in the 
experiments on SH3 \cite{Riddle}.
The computational scheme for the clustering is stochastic  in nature 
and does not permit a single entry group due to technical reasons associated
with the computation.

Here, we implement a much simpler and deterministic clustering scheme
which is based on considering the "distances" between the
amino acids.  The groupings we get are different from those obtained by
Wang and Wang \cite{Wang}.   
We have carried out detailed tests of two characteristics
of real proteins -- thermodynamic stability and kinetic accessibility --
on well-designed sequences within the context of a lattice 
model in three dimensions.
When the individual amino acids are substituted by the representatives of their
groups, the thermodynamic stability test is passed successfully both by
the Wang-Wang and our schemes.  However, there is a qualitative difference
in the test results on rapid folding into the native state, with the 
Wang-Wang scheme 
performing significantly worse than our physically-based approach.
A straightforward eigenvalue analysis of the kind carried out 
by Li et al. \cite{Tang}
and Chan \cite{Chan} lead to results in perfect accord with our clustering
scheme.  Furthermore, when one works with the bare MJ 
matrix \cite{Chan} (without
subtracting off the mean value as done by Li et al. \cite{Tang}),
one obtains the simple result that the MJ matrix may be represented as
\begin{equation}
M^{1}_{ij} \;=\; -\lambda _1 v_i v_j \;\;,
\end{equation}
where $\lambda _1$=69.68 and $v$ has rank-ordered components
0.333, 0.332, 0.309, 0.282, 0.276, 0.275, 0.255, 0.249, 0.198, 0.196,
0.177, 0.171, 0.169, 0.163, 0.161, 0.158, 0.152, 0.144, 0.144, 0.125
in the order LFIMVWCYHATGPRQSNEDK respectively.
The representation given in eq. 1 is similar in spirit to the
reprezentation \cite{Tang} in terms of ``charges", $q_i$, 
which yields $M'_{ij}=C_0 + C_1(q_i+q_j) + C_2q_iq_j$. Eq. (1) results
in the same level of accuracy (an average error of about 6\%),
even though it involves only one constant, $\lambda_1$, instead of three,
$C_0$, $C_1$, and $C_2$.

The interactions
of a given amino acid $i$ with each of the 20 amino acids
form a vector with 20 components.  There are 20 such vectors and the
Euclidean distance, $R_{ij}$, between amino acids $i$ and $j$
is defined through $R^2_{ij}=\sum _k (M_{ik}-M_{jk})^2$.
Similarly, the Manhattan distance involves the sum of the absolute 
values of $M_{ik}-M_{jk}$.  $R_{ij}$ is a measure of the fidelity
of substitution of one amino acid by the other.
As in the construction of optimal paths in a strongly disordered
medium \cite{Cieplak}, we select two
amino acids that are separated by the shortest distance and combine
them into one group. The effective couplings of a group with other groups
or individual amino acids  or indeed with itself is simply obtained as
an arithmetic average over the individual amino acid interactions.
More generally, a weighted average would be appropriate if the
weights, which could depend on the frequency of occurrence of 
a given amino acid and other factors, were known.
The procedure is now iterated
resulting in fewer and fewer groups. The leader 
of the group is determined as the amino acid whose original couplings
deviate the least from the effective couplings associated with the group.
For a two-member group the choice of a leader is ambiguous.
At each stage, the advancement of clustering is characterized by
measuring the smallest distance between the remaining 
individuals or groups, $R_{min}$.
Figure 2 shows the behavior $R_{min}$ as a function of $N_a$, the number of
groups at any stage of the iteration process.
For either distance metric, the number $N_c = 5$ is a special number 
of groups, below which $R_{min}$
goes up strongly.  This sharp increase is a direct reflection of the
forced clustering of incompatible amino acids.

Our analysis can also be used to probe the nature of other theoretically
derived potentials of interaction between amino acids.  For example, for
the matrix of interactions proposed by 
Kolinski et al. \cite{KGS}, the low $N_a$ growth is much weaker
but $N_c$ is still around 5.  For the matrix introduced by Betancourt and 
Thirumalai \cite{Betancourt},  $R_{min}$ varies very weakly with $N_a$
and the sharp increase takes place at $N_c$ equal to 3.

Our results for clustering of the Miyazawa-Jernigan amino acids
are shown in Figure 1. Our clustering into five groups
separates the single hydrophobic group of Wang and Wang into two groups
and it isolates K as a single element group. The membership of the five groups
does not depend on the distance metric, even though different
amino acids are selected as ``leaders", or the most characteristic 
representatives of the groups.
The Euclidean choice proposes FVASK as the best
whereas the choice of the Manhattan metric yields LWASK.
These groupings are consistent with the experimental results
except that Riddle et al. \cite{Riddle} take two choices, E and G, from the
TGPRQSNED group and no choice from the second hydrophobic group MVWCY.
It is heartening that the ultimate division into hydrophobic and polar groups
is quite robust and yields the same group members in both schemes and for any
distance metric.
In order to test the five group clustering,
we have considered a 27-monomer lattice model 
with the Miyazawa-Jernigan
couplings. The native states are maximally compact, i.e. they fit
the $3\times 3 \times 3$ lattice. We generate a bank of 
94 well designed sequences. The first step in the
design procedure \cite{Shakhnovich} involved the generation of 
a random sequence with a uniform composition. 
The second step was to perform random permutations in the 
placement of the amonoacids and to select only those sequences which have
energy gaps (the energy difference between the two lowest energy
maximally compact conformations) larger than 6.1 
(in units of the MJ interaction matrix).
This bank of sequences was then tested against substitution of the
amino acids by the leaders corresponding to the five-group division.
For both the distance-based and mismatch based schemes we find
identical performance -- 95\% of the resulting effective sequences continue
to have nondegenerate native states which coincide with the
native state conformations of the original sequences. 
(A lower performance rate of 80\%, quoted by Wang and Wang \cite{Wang}, 
is obtained on considering
less stringently designed sequences.)

A more stringent test is provided by the kinetics. 
Figure 3 shows the folding behavior of four of the well-designed sequences.
The first passage time in 21 runs starting from randomly chosen 
unfolded conformations are shown for all cases which have a folding time
less than $5 \times 10^8 \tau_0$.
The dynamics are based on a Monte Carlo process which satisfies 
detailed balance \cite{Henkel}.
The folding is studied at the
folding transition temperature, $T_f$, at which 
the probability of being in the native 
conformation equals $\frac{1}{2}$. This temperature was located
using a long unfolding Monte Carlo process, which explores the entire 
space of conformations and is not restricted merely to maximally
compact conformations.
The distance based substitution
either raises $T_f$ or leaves it intact and preserves the range of 
values of the folding times.
On the other hand, the mismatch
based substitution lowers all four values of $T_f$ (some of them significantly)
and extends folding times
substantially.   The simple origin of this failure may be traced to
the fact that $K$ which is a relatively inert amino acid is chosen as 
one of the leaders in the mismatch scheme and sequences with many amino
acids belonging to the group with $K$ are all substituted by an
innocuous representative.

The clustering schemes illustrated in Figure 1 correspond to the partitioning
of a string of all amino acids into segments provided that  the
amino acids are first arranged into a particular order. The order 
corresponding to the mismatch based scheme is almost the same as 
shown in Figure 1 for the distance based scheme except that K is forced
to be placed earlier along the string. What is it that determines
this optimal order?
Following Li et al. \cite{Tang} and Chan \cite{Chan}, 
we address this question through an eigenanalysis of the MJ matrix.
As shown by Chan, the MJ matrix is clearly dominated 
by one mode and lends itself to an exceedingly simple representation.
Figure 4 shows that this one mode description of the MJ matrix
is fairly accurate (the largest errors are located at the smallest
magnitude entries). In fact, it is as accurate as the two mode 
description of Li et al. \cite{Tang}. Their two mode picture
emerges because of the subtraction of the mean value from each matrix element.
This mean value has physical significance because it determines
the degree of compactness and indeed the degree of aggregation
of a protein \cite{Giugli}.
It is interesting to note that the dominant eigenvector varies across
the amino acids in a smooth way except near the
transition from Y to H which marks a transition from the
hydrophobic to polar groups.
The order in which the amino acids are plotted on the x axis in Figure 4
and in (most of) Figure 1 is thus determined by the weight with which
a particular amino acid contributes to the dominant eigenvector.
In summary, we have shown that a simple distance-based
scheme of clustering leads to a powerful and simple representation 
of groupings of amino acids making up a protein.  Detailed
tests within the context of a MJ matrix  and a lattice model show that both
the thermodynamic stability and the folding kinetics of protein-like
sequences are preserved by the substitution of the full 20 amino acid
alphabet by merely 5 groups.  Our scheme is found to be in good
physical accord with that obtained using eigenanalysis and does not
have the defects associated with a more complex mismatch based scheme.

We are grateful to Ruxandra Dima for fruitful discussions.
This work was supported 
by KBN (Poland - grant 2P03B-146-18),        
MURST (Italy), NASA, IGERT-NSF and the
Petroleum Research Fund administered by the American Chemical
Society.

\vspace*{-13cm}

\begin{figure}
\epsfxsize=3.0in
\vspace*{0.5cm}
\centerline{\epsffile{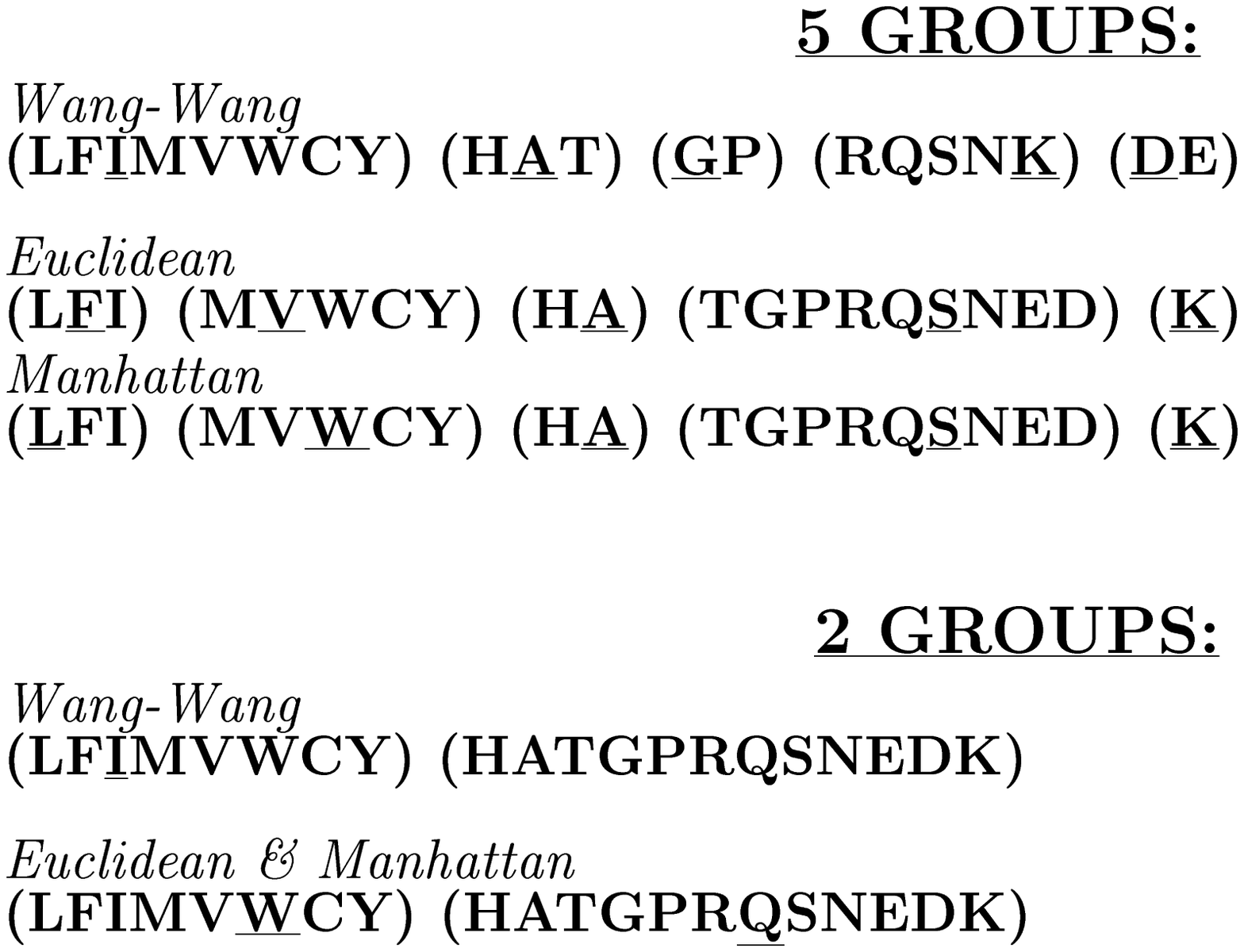}}
\vspace*{-1.5cm}
\caption{Five- and two-group clustering of amino acids interacting
through the Miyazawa-Jernigan matrix. The mismatch-based results are 
shown at the top in each level of clustering and the remaining
entries are distance-based. 
The underlined aminoacids are the group representatives.
}
\end{figure}

\begin{figure}
\epsfxsize=2.8in
\centerline{\epsffile{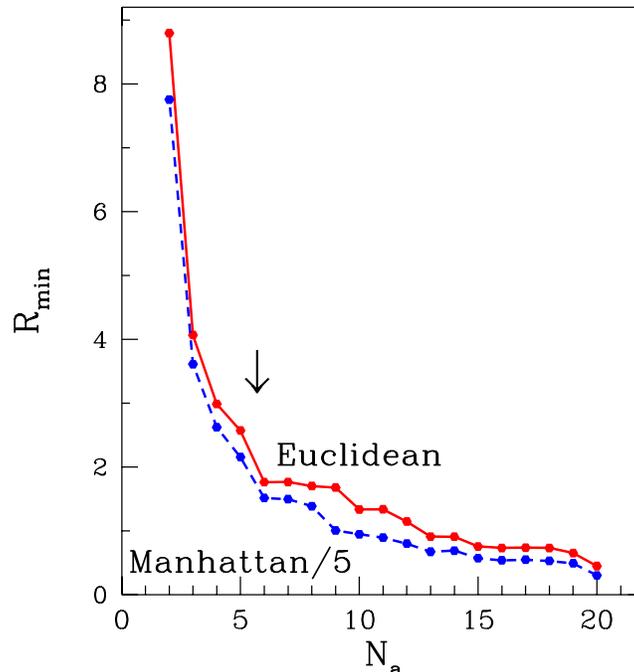}}
\vspace*{1.8cm}
\caption{Minimal distance between the groups of amino acids as a function
of the number of groups present. The arrow indicates a stage in which
one is left with five groups.
}
\end{figure}

\begin{figure}
\epsfxsize=2.9in
\centerline{\epsffile{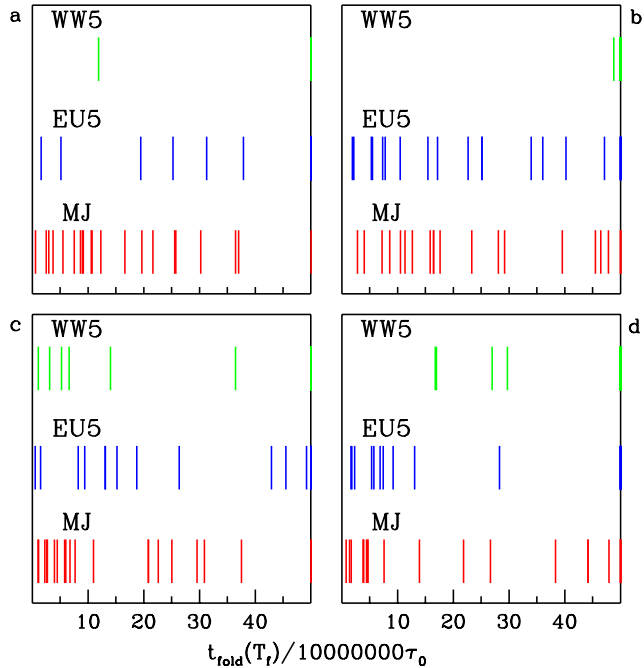}}
\vspace*{0.5cm}
\caption{ The bottom set of lines (marked as MJ) in each panel shows the
folding times for a well-designed sequence using the original MJ interaction
matrix:
FLISKQINECKEFDLRLGEHSTSCQVK (a, top left),
QQLASQHLLTRWNWNMNHNPRSIDFQF (b, top right), 
DVECEVSVQWIQKHFRLTFTIFMYEAD (c, bottom left), and
HMKLRSFPSIQVEMRVFDFRLTFIIRA (d, bottom right) at 
their folding transition temperatures.
The folding transition temperatures are 1.48, 1.58, 1.68 and 1.75 respectively
in units of the MJ interaction matrix.
In each case, we consider 21 folding trajectories and
show the folding times only when they are shorter than  500 000 000 $\tau _0$,
where $\tau _0$ is a microscopic time scale that takes into account the
maximum number of moves from a any of the available conformations.
The center set of lines (marked as EU5)
correspond to sequences in which the Euclidean distance-based
substitution of Figure 1 is implemented. On substitution, the values of
$T_f$ become 1.48, 1.68, 1.58 and 1.79 respectively. Note the slight
increase in the thermodynamic stability in two of the cases.
The top set of lines (marked as WW5) correspond 
to substitutions recommended by Wang and Wang.
Only few of the folding attempts occur within the window of 
500 000 000 $\tau _0$
which indicates a significant deterioration in the folding kinetics.
The folding transition temperatures are reduced
(substantially in two cases). They become
0.31, 1.28, 1.53 and 1.48 respectively.
} 
\end{figure}

\begin{figure} 
\epsfxsize=2.9in 
\centerline{\epsffile{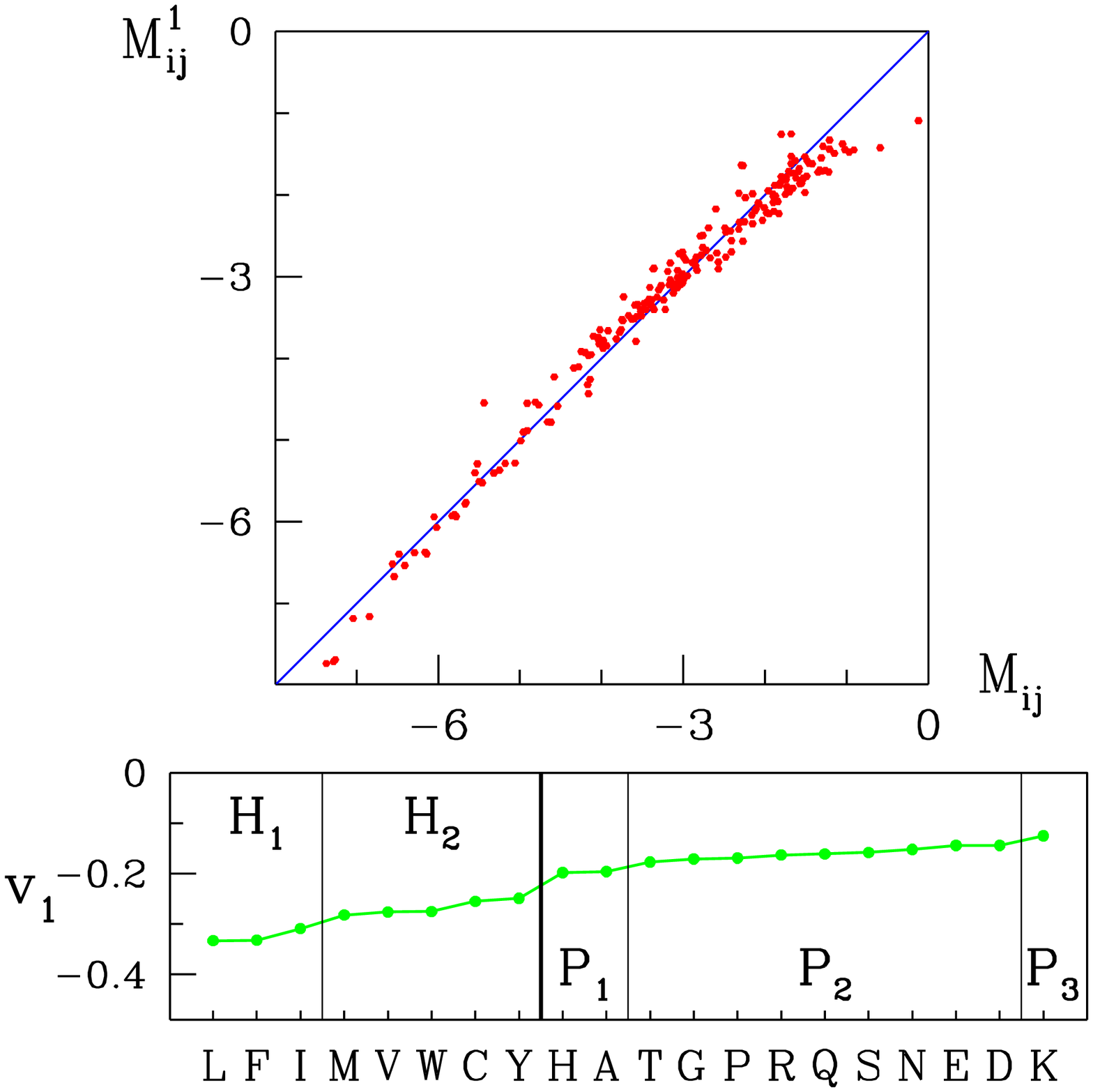}} 
\vspace*{1.5cm} 
\caption{Top: The approximation $M^1$ to the MJ matrix, given by eq. 1, 
is plotted versus the true matrix elements.
Bottom: The  eigenvector corresponding to the largest 
eigenvalue of the MJ matrix. The vertical
lines indicate partitioning into five groups of Figure 1 as determined by the
distance based method. The symbols H$_1$ and H$_2$ indicate two
hydrophobic groups. P$_1$, P$_2$, and P$_3$ indicate three polar groups.
}
\end{figure}


\vspace{0.5cm}

\begin{references}

\bibitem{Miyazawa}
S. Miyazawa and R. L. Jernigan, 
J. Mol. Biol. {\bf 256}, 623-644 (1996).

\bibitem{Kauzmann}
W. Kauzmann, in {\it The Mechanism of Enzyme Action}, eds. W. D. McElroy
and B. Glass, 70-120 (Johns Hopkins Press, Baltimore, 1954);
W. Kauzmann, Adv. Protein Chem. {\bf 14}, 1-63 (1959).

\bibitem{Kamtekar}
S. Kamtekar, J. M. Schiffer, H. Xiong, J. M. Babik, and M. Hecht,
Science {\bf 262}, 1680-1685 (1993).

\bibitem{Dill}
K. F. Lau and K. A. Dill, 
Macromolecules {\bf 22}, 3986-3997 (1989);
K. A. Dill, S. Bromberg, S. Yue, K. Fiebig, K. M. Yee, D. P. Thomas,
and H. S. Chan, Protein Sci. {\bf 4}, 561 (1995).

\bibitem{Regan}
L. Regan and W. E. DeGrado,
Science {\bf 241}, 976-978 (1988).

\bibitem{Riddle}
D. S. Riddle, J. V. Santiago, S. T. Bray-Hall, N. Doshi, V. P. Grantcharova,
Q. Yi, and D. Baker,
Nature Struct. Biol. {\bf 4}, 805-809 (1997).

\bibitem{Wolynes}
P. G. Wolynes,
Nature Struct. Biol. {\bf 4}, 871-874 (1997).

\bibitem{Wang}
J. Wang and W. Wang,
Nature Struct. Biol. {\bf 6}, 1033-1038 (1999).


\bibitem{Tang}
H. Li, C. Tang, and N. S. Wingreen, 
Phys. Rev. Lett. {\bf 79}, 765-768
(1997).

\bibitem{Chan}
H. S. Chan,
Nature Struct. Biol. {\bf 6}, 994-996 (1999).

\bibitem{Cieplak}
M. Cieplak, A. Maritan, and J. R. Banavar,
Phys. Rev. Lett. {\bf 72}, 2320-2323 (1994).

\bibitem{KGS}
A. Kolinski, A. Godzik, and J. Skolnick, J. Chem. Phys. {\bf 98},
7420-7433 (1993).

\bibitem{Betancourt}
M. R. Betancourt and D. Thirumalai,
Protein Science {\bf 8}, 361-369 (1999)

\bibitem{Shakhnovich}
E. I. Shakhnovich and A. M. Gutin, 
Proc. Natl. Acad. Sci. {\bf 90} 7195-7199 (1993) 

\bibitem{Henkel}
M. Cieplak, M. Henkel, J. Karbowski, and J. R. Banavar,
Phys. Rev. Lett. {\bf 80}, 3654-3657 (1998);
M. Henkel, and J. R. Banavar, J. Cond. Mat. {\bf 2}, 369-378 (1999)

\bibitem{Giugli}
G. Giugliarelli, C. Micheletti, J. R. Banavar, A. Maritan, J. Chem. Phys.
{\bf 113}, 5072 (2000)




\end{references}
\end{document}